\documentclass[conference]{IEEEtran}
\usepackage{amsfonts}
\usepackage{amsfonts}

\usepackage{amsmath}
\usepackage{amssymb}
\usepackage[dvips]{graphicx}
\usepackage{epsfig}
\usepackage{hyperref}

\usepackage{amsmath}
\usepackage{amssymb}
\usepackage[dvips]{graphicx}
\usepackage{epsfig}

\newcommand{\tsnr}{{\text{\footnotesize{SNR}}}}
\newcommand{\tmin}{\text{min}}
\newcommand{\E}{\mathbb{E}}
\newcommand{\e}{\mathcal{E}}

\newcommand{\R}{{\sf{R}}}
\newcommand{\Pb}{\bar{P}}
\newcommand{\sta}{{\alpha^{\ast}_{\text{opt}}}}
\newcommand{\alphao}{\alpha_{\text{opt}}}
\newcommand{\ro}{r_{\text{opt}}}

\newcommand{\figsize}{0.37}
\newcommand{\rhoo}{\rho_{\text{opt}}}
\newcommand{\tsnref}{\tsnr_{\text{eff}}}
\newcommand{\tsnrefo}{\tsnr_{\text{eff,opt}}}

\newtheorem{Lem}{Theorem}

\begin{document}

%
\title{Energy Efficiency of Fixed-Rate Wireless Transmissions under Queueing Constraints and Channel Uncertainty}



%
\author{\authorblockN{Deli Qiao, Mustafa Cenk Gursoy, and Senem
Velipasalar}
\authorblockA{Department of Electrical Engineering\\
University of Nebraska-Lincoln, Lincoln, NE 68588\\ Email:
qdl726@bigred.unl.edu, gursoy@engr.unl.edu, velipasa@engr.unl.edu}}


\maketitle

\begin{abstract}\footnote{This work was supported by the National Science Foundation under Grants CCF - 0546384 (CAREER) and CNS - 0834753.}
Energy efficiency of fixed-rate transmissions is studied in the presence of queueing constraints and channel uncertainty. It is assumed that neither the transmitter nor the
receiver has channel side information prior to transmission. The channel coefficients are estimated at the receiver via minimum mean-square-error (MMSE) estimation with the aid of training symbols. It is further assumed that the system operates under statistical queueing constraints in the form of limitations on buffer violation probabilities. The optimal fraction of of power allocated to training is
identified. Spectral efficiency--bit energy
tradeoff is analyzed in the low-power and wideband regimes by
employing the effective capacity formulation.  In particular, it is shown that the bit energy increases without bound in the low-power regime as the average power vanishes. On the other hand, it is proven that the bit energy diminishes to its minimum value in the wideband regime as the available bandwidth increases. For this case, expressions for the minimum bit energy and wideband slope are derived.  Overall, energy costs of channel uncertainty and queueing constraints are identified.
\end{abstract}
\section{Introduction}

The two key characteristics of wireless communications that most greatly impact system design and performance are 1) the
randomly-varying channel conditions and 2) limited energy resources. In wireless systems, the power of the received signal fluctuates
randomly over time due to mobility, changing environment, and
multipath fading.
These random changes in the received signal strength lead to
variations in the instantaneous data rates that can be supported by
the channel \cite{book}. In addition, mobile wireless systems can only be
equipped with limited energy resources, and hence energy efficient
operation is a crucial requirement in most cases.

To measure and compare the energy efficiencies of different systems
and transmission schemes, one can choose as a metric the energy
required to reliably send one bit of information.
Information-theoretic studies show that energy-per-bit requirement
is generally minimized, and hence the energy efficiency is
maximized, if the system operates at low signal-to-noise ratio
($\tsnr$) levels and hence in the low-power or wideband regimes.
Recently, Verd\'u in \cite{sergio} has determined the minimum bit
energy required for reliable communication over a general class of
channels, and studied of the spectral efficiency--bit energy
tradeoff in the wideband regime while also providing novel tools
that are useful
for analysis at low $\tsnr$s. 

In many wireless communication systems, in addition to
energy-efficient operation, satisfying certain quality of service
(QoS) requirements is of paramount importance in providing
acceptable performance and quality. For instance, in voice over IP
(VoIP), interactive-video (e.g,. videoconferencing), and
streaming-video applications in wireless systems, latency is a key
QoS metric and should not exceed certain levels \cite{Szigeti}. On
the other hand, wireless channels, as described above, are
characterized by random changes in the channel, and such volatile
conditions present significant challenges in providing QoS
guarantees. In most cases, statistical, rather than deterministic,
QoS assurances can be given.


In summary, it is vital for an important class of wireless systems
to operate efficiently while also satisfying QoS requirements (e.g.,
latency, buffer violation probability). Information theory provides
the ultimate performance limits and identifies the most efficient
use of resources. However, information-theoretic studies and Shannon
capacity formulation generally do not address delay and quality of
service (QoS) constraints \cite{Ephremides}. Recently, Wu and Negi
in \cite{dapeng} defined the effective capacity as the maximum
constant arrival rate that a given time-varying service process can
support while providing statistical QoS guarantees. Effective
capacity formulation uses the large deviations theory and
incorporates the statistical queueing constraints by capturing the rate
of decay of the buffer occupancy probability for large queue
lengths. The analysis and application of effective capacity in
various settings has attracted much interest recently (see e.g.,
\cite{dapeng}--\cite{fixed} and references therein). 

In this paper, we study the energy efficiency in the presence of queueing constraints and channel uncertainty. We assume that the channel is not known by the transmitter and receiver prior to transmission, and is estimated imperfectly by the receiver through training. In our model, we incorporate statistical queueing constraints by employing the effective capacity formulation which provides the maximum throughput under limitations on buffer violation probabilities for large buffer sizes. Since the transmitter is assumed to not know the channel, fixed-rate transmission is considered.

\section{System Model}
We consider a point-to-point
wireless link in which there is one source and one destination. It
is assumed that the source generates data sequences which are
divided into frames of duration $T$. These data frames are initially
stored in the buffer before they are transmitted over the wireless
channel. The discrete-time channel input-output relation in the
$i^{\text{th}}$ symbol duration is given by
\begin{gather} \label{eq:model}
y[i] = h[i] x[i] + n[i] \quad i = 1,2,\ldots.
\end{gather}
where $x[i]$ and $y[i]$ denote the complex-valued channel input and
output, respectively. We assume that the bandwidth available in the
system is $B$ and the channel input is subject to the following
average energy constraint: $\E\{|x[i]|^2\}\le \Pb / B$ for all $i$.
Since the bandwidth is $B$, symbol rate is assumed to be $B$ complex
symbols per second, indicating that the average power of the system
is constrained by $\Pb$. Above in (\ref{eq:model}), $n[i]$ is a
zero-mean, circularly symmetric, complex Gaussian random variable
with variance $\E\{|n[i]|^2\} = N_0$. The additive Gaussian noise
samples $\{n[i]\}$ are assumed to form an independent and
identically distributed (i.i.d.) sequence. Finally, $h[i]$, which denotes
the channel fading coefficient, is assumed to be a zero-mean Gaussian random variable
with variance $E\{|h|^2\} = \gamma$. We further assume that the fading coefficients stay constant during the frame duration of $T$ seconds and have independent realizations for each frame. Hence, we basically consider a block-fading channel model. We also note that neither the transmitter not the receiver has channel side information prior to transmission.

The system operates in two phases: training phase and data
transmission phase. In the training phase, known pilot symbols are
transmitted to enable the receiver to estimate the channel
conditions, albeit imperfectly. We assume that
minimum mean-square-error (MMSE) estimation is employed at the
receiver to estimate the channel coefficient $h[i]$. Since the
MMSE estimate depends only on the training energy and not on the
training duration \cite{training}, it can be easily seen that transmission of a
single pilot at every $T$ seconds is optimal. Note that in every
frame duration of $T$ seconds, we have $TB$ symbols and the overall
available energy is $\Pb T$. We now assume that each frame consists
of a pilot symbol and $TB - 1$ data symbols. The energies of the
pilot and data symbols are
\begin{equation}\label{eq:trainpower}
\e_t=\rho \Pb T, \quad\text{and}\quad \e_s=\frac{(1-\rho)\Pb
T}{TB-1},
\end{equation}
respectively, where $\rho$ is the fraction of total energy allocated
to training. Note that the data symbol energy $\e_s$ is obtained by
uniformly allocating the remaining energy among the data symbols.

In the training phase, the
receiver obtains the MMSE estimate $\hat{h}$ which is a circularly
symmetric, complex, Gaussian random variable with mean zero and
variance $\frac{\gamma^2 \e_t}{\gamma \e_t + N_0}$, i.e.,
$\hat{h} \sim \mathcal{CN} \left( 0, \frac{\gamma^2 \e_t}{\gamma
\e_t + N_0} \right)$\cite{gursoy}.
Now, the
channel fading coefficient $h$ can be expressed as
$h=\hat{h}+\tilde{h}$
where $\tilde{h}$ is the estimate error and $\tilde{h}\sim\mathcal
{CN}(0,\frac{\gamma N_0}{\gamma \e_t+N_0})$. Consequently, in the data transmission phase, the
channel input-output relation becomes
\begin{gather} \label{eq:impmodel}
y[i] = \hat{h}[i] x[i] + \tilde{h}[i] x[i] + n[i] \quad i =
1,2,\ldots.
\end{gather}
Since finding the capacity of the channel in (\ref{eq:impmodel}) is
a difficult task\footnote{In \cite{gursoy}, the capacity of
training-based transmissions under input peak power constraints is
shown to be achieved by an $\tsnr$-dependent, discrete distribution
with a finite number of mass points. In such cases, no closed-form
expression for the capacity exists, and capacity values need to be
obtained through numerical computations.}, a capacity lower bound is
generally obtained by considering the estimate error $\tilde{h}$ as
another source of Gaussian noise and treating $\tilde{h}[i] x[i] +
n[i]$ as Gaussian distributed noise uncorrelated from the input.
Now, the new noise variance is $\E\{|\tilde{h}[i] x[i] + n[i]|^2\} =
\sigma_{\tilde{h}}^2 \e_s + N_0$ where $\sigma_{\tilde{h}}^2 =
\E\{|\tilde{h}|^2\} = \frac{\gamma N_0}{\gamma \e_t+N_0}$ is the
variance of the estimate error. Under these assumptions, a lower
bound on the instantaneous capacity is given by \cite{training},
\cite{gursoy}
\begin{align}
C_L&=\frac{TB-1}{T}\log_2\left(1+ \frac{\e_s}{\sigma_{\tilde{h}}^2
\e_s + N_0} |\hat{h}|^2\right)\\
& =\frac{TB-1}{T} \log_2\left(1+\tsnref |w|^2\right) \text{ bits/s}
\label{eq:traincap2}
\end{align}
where effective $\tsnr$ is
\begin{equation}\label{eq:trainsnr}
\tsnref=\frac{\e_s \sigma_{\hat{h}}^2}{\sigma_{\tilde{h}}^2 \e_s +
N_0},
\end{equation}
and $\sigma^2_{\hat{h}} = \E\{|\hat{h}|^2\} = \frac{\gamma^2
\e_t}{\gamma \e_t + N_0}$ is the variance of estimate $\hat{h}$.
Note that the expression in (\ref{eq:traincap2}) is
obtained by defining $\hat{h} = \sigma_{\hat{h}} w$ where $w$ is a
standard complex Gaussian random variable with zero mean and unit variance,
i.e., $w\sim\mathcal {CN}(0,1)$.

Since Gaussian is the worst uncorrelated noise \cite{training}, the
above-mentioned assumptions lead to a pessimistic model and   the
rate expression in (\ref{eq:traincap2}) is a lower bound to the
capacity of the true channel (\ref{eq:impmodel}). On the other hand,
$C_L$ is a good measure of the rates achieved in communication
systems that operate as if the channel estimate were perfect (i.e.,
in systems where Gaussian codebooks designed for known channels are
used, and scaled nearest neighbor decoding is employed at the
receiver) \cite{lapidoth}.

\begin{figure}
\begin{center}
\includegraphics[width=\figsize\textwidth]{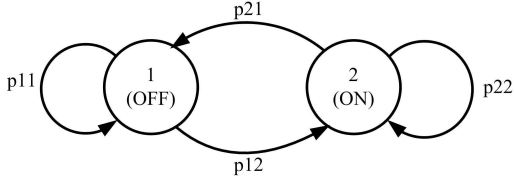}
\caption{ON-OFF state transition model.}\label{fig:00}
\end{center}
\end{figure}

Henceforth, we base our analysis on $C_L$ to understand the impact
of the imperfect channel estimate. Since the transmitter is unaware of the
channel conditions, it is assumed that information is transmitted at a fixed rate of
$r$ bits/s. When $r < C_L$, the channel is considered to be in the
ON state and reliable communication is achieved at this rate. If, on
the other hand, $r \ge C_L$, we assume that outage occurs. In this case, channel is in
the OFF state and reliable communication at the rate of $r$ bits/s
cannot be attained. Hence, effective data rate is zero and
information has to be resent.
Fig. \ref{fig:00} depicts the two-state transmission model together
with the transition probabilities.
Under the block fading assumption, it can be easily seen that the transition probabilities are given by
\begin{align}
p_{11}&=p_{21}= P\{r \ge C_L\} =P\{|w|^2 \le \alpha\} \\
p_{22}&=p_{12}= P\{r < C_L\}= P\{|w|^2 > \alpha\}
\end{align}
where
\begin{equation}\label{eq:trainthresh}
\alpha=\frac{2^{\frac{rT}{TB-1}}-1}{\tsnref},
\end{equation}
and $|w|^2$ is an exponential random variable with mean $1$, and
hence, $P\{|w|^2 > \alpha\} = e^{-\alpha}$.

%

\section{Effective Capacity and Spectral Efficiency--Bit Energy Tradeoff}

In \cite{dapeng}, Wu and Negi defined the effective capacity as the
maximum constant arrival rate that a given service process can
support in order to guarantee a statistical QoS requirement
specified by the QoS exponent $\theta$ \footnote{For time-varying
arrival rates, effective capacity specifies the effective bandwidth
of the arrival process that can be supported by the channel.}. If we
define $Q$ as the stationary queue length, then $\theta$ is the
decay rate of the tail distribution of the queue length $Q$:
\begin{equation}
\lim_{q \to \infty} \frac{\log P(Q \ge q)}{q} = -\theta.
\end{equation}
Therefore, for large $q_{\max}$, we have the following approximation
for the buffer violation probability: $P(Q \ge q_{\max}) \approx
e^{-\theta q_{max}}$. Hence, while larger $\theta$ corresponds to
more strict QoS constraints, smaller $\theta$ implies looser QoS
guarantees. Similarly, if $D$ denotes the steady-state delay
experienced in the buffer, then $P(D \ge d_{\max}) \approx
e^{-\theta \delta d_{\max}}$ for large $d_{\max}$, where $\delta$ is
determined by the arrival and service processes
\cite{tangzhangcross2}.

The effective capacity is given by
\begin{gather}
-\frac{\Lambda(-\theta)}{\theta}=-\lim_{t\rightarrow\infty}\frac{1}{\theta
t}\log_e{\mathbb{E}\{e^{-\theta S[t]}\}}
\end{gather}
where $S[t] = \sum_{i=1}^{t}R[i]$ is the time-accumulated service
process and $\{R[i], i=1,2,\ldots\}$ denote the discrete-time
stationary and ergodic stochastic service process. Note that in the
model we consider, $R[i] = rT \text{ or } 0$ depending on the
channel state being ON or OFF. In \cite{cschang}, it is shown that
for such an ON-OFF model, we have
%
\begin{align}
\frac{\Lambda(\theta)}{\theta}&=\frac{1}{\theta}\log_e\Big(\frac{1}{2}\Big(p_{11}+p_{22}e^{\theta
Tr} \nonumber \\
&\hspace{-.08cm}+\sqrt{(p_{11}+p_{22}e^{\theta
Tr})^2+4(p_{11}+p_{22}-1)e^{\theta Tr}} \Big)\Big)
\end{align}
Note that $p_{11}+p_{22}=1$ in our  model. Then, for a given QoS
delay constraint $\theta$, the effective capacity normalized by the
frame duration $T$ and bandwidth $B$, or equivalently spectral
efficiency in bits/s/Hz, becomes
\begin{align}
\R_E(\tsnr,\theta)&=\max_{\substack{r\geq0 \\ 0\leq
\rho\leq1}}{-\frac{1}{
TB}\frac{\Lambda(-\theta)}{\theta}} \quad \text{bits/s/Hz}\\
&=\max_{\substack{r\geq0 \\ 0\leq \rho\leq1}}{-\frac{1}{\theta
TB}\log_e\big(p_{11}+p_{22}e^{-\theta
Tr}\big)}\\
&=\max_{\substack{r\geq0 \\ 0\leq \rho\leq1}}{-\frac{1}{\theta
TB}\log_e\big(1-P(|w|^2>\alpha)(1-e^{-\theta
Tr})\big)}  \label{eq:trainopti}
\end{align}
\begin{align}
&=-\frac{1}{\theta TB}\log_e\big(1-P(|w|^2>\alphao)(1-e^{-\theta
T\ro})\big)  \label{eq:trainopti2}.
\end{align}
Note that $\R_E$ is obtained by optimizing both the fixed
transmission rate $r$ and the fraction of power allocated to
training, $\rho$. In the optimization result (\ref{eq:trainopti2}),
$\ro$ and $\alphao$ are the optimal values of $r$ and $\alpha$,
respectively. It can easily be seen that
\begin{align}
\R_E(\tsnr,0) &= \lim_{\theta \to 0} \R_E(\tsnr,\theta) \\
 &=
\max_{r\ge 0} \,\,\frac{r}{B} \, P\left\{|w|^2 >
\frac{2^{\frac{rT}{TB-1}}-1}{\tsnref}\right\}.
\end{align}
Hence, as the QoS requirements relax, the maximum constant arrival
rate approaches the average transmission rate. On the other hand, for $\theta > 0$, $\R_E <
\frac{1}{B} \max_{r\ge 0} r P(|w|^2 >\alpha)$ in order to avoid
violations of buffer constraints.

In this paper, we focus on the energy efficiency of wireless
transmissions under the aforementioned statistical queueing constraints. Since energy efficient operation generally requires operation   at low-$\tsnr$ levels, our analysis throughout the paper is carried out
in the low-$\tsnr$ regime. In
this regime, the tradeoff between the normalized effective capacity
(i.e, spectral efficiency) $\R_E$ and bit energy $\frac{E_b}{N_0} =
\frac{\tsnr}{\R_E(\tsnr)}$ is a key tradeoff in understanding the
energy efficiency, and is characterized by the bit energy at zero
spectral efficiency and wideband slope provided, respectively, by
\begin{equation}\label{eq:ebresult}
\frac{E_b}{N_0}\bigg|_{\R = 0}
= \frac{1}{\dot{\R}_{E}(0)} \text{ and
} \mathcal{S}_0=-\frac{2(\dot{\R_E}(0))^2}{\ddot{\R_E}(0)}\log_e{2}
\end{equation}
where $\dot{\R}_E(0)$ and $\ddot{\R}_E(0)$ are the first and second
derivatives with respect to $\tsnr$, respectively, of the function $\R_E(\tsnr)$ at zero $\tsnr$ \cite{sergio}.
$\frac{E_b}{N_0}\Big|_{\R=0}$ and $\mathcal{S}_0$ provide a linear
approximation of the spectral efficiency curve at low spectral
efficiencies.

\section{Optimal Power Allocation for Training}

In this section, we investigate the optimization problem
in (\ref{eq:trainopti}). In particular, we identify the optimal fraction of power that needs to be allocated to training while satisfying statistical buffer constraints.
\begin{Lem} \label{theo:optrho}
At a given $\tsnr$ level, the optimal fraction of power $\rhoo$ that
solves (\ref{eq:trainopti}) does not depend on the QoS exponent
$\theta$ and the transmission rate $r$, and is given by
\begin{equation}\label{eq:optrho}
\rhoo=\sqrt{\eta(\eta+1)}-\eta
\end{equation}
where $\eta=\frac{\gamma TB\tsnr+TB-1}{\gamma TB(TB-2)\tsnr}$ and
$\tsnr = \frac{\Pb}{N_0B}$.
\end{Lem}
\emph{Proof:} From (\ref{eq:trainopti}) and the definition of
$\alpha$ in (\ref{eq:trainthresh}), we can easily see that for fixed
$r$, the only term in (\ref{eq:trainopti}) that depends on $\rho$ is
$\alpha$. Moreover, $\alpha$ has this dependency through $\tsnref$.
Therefore, $\rhoo$ that maximizes the objective function in
(\ref{eq:trainopti}) can be found by minimizing $\alpha$, or
equivalently maximizing $\tsnref$. Substituting the definitions in
(\ref{eq:trainpower}) and the expressions for $\sigma_{\hat{h}}^2$
and $\sigma_{\tilde{h}}^2$ into (\ref{eq:trainsnr}), we have
\begin{align}
\tsnref&=\frac{\e_s \sigma_{\hat{h}}^2}{\sigma_{\tilde{h}}^2 \e_s +
N_0} \\
&= \frac{\rho(1-\rho)\gamma^2T^2B^2\tsnr^2}{\rho \gamma
TB(TB-2)\tsnr+\gamma TB\tsnr+TB-1}\label{eq:trainsnref}
\end{align}
where $\tsnr=\frac{\Pb}{N_0 B}$. Evaluating the derivative of
$\tsnref$ with respect to $\rho$ and making it equal to zero leads
to the expression in (\ref{eq:optrho}). Clearly, $\rhoo$ is
independent of $\theta$ and $r$.

Above, we have implicitly assumed that the maximization is performed
with respect to first $\rho$ and then $r$. However, the result will
not alter if the order of the maximization is changed. Note that the
objective function in (\ref{eq:trainopti})
\begin{align}
g(\tsnref,r)&= - \frac{1}{\theta
TB}\log_e\bigg(1-P\left(|w|^2>\frac{2^{\frac{rT}{TB-1}}-1}{\tsnref}\right)\nonumber\\
&\phantom{- \frac{1}{\theta TB}\log_e\bigg(1-}\times(1-e^{-\theta
Tr})\bigg)
\end{align}
is a monotonically increasing function of $\tsnref$ for all $r$. It
can be easily verified that maximization does not affect the
monotonicity of $g$, and hence $\max_{r \ge 0} g(\tsnref,r)$ is
still a monotonically increasing function of $\tsnref$. Therefore,
in the outer maximization with respect to $\rho$, the choice of
$\rho$ that maximizes $\tsnref$ will also maximize $\max_{r \ge 0}
g(\tsnref,r)$, and the optimal value of $\rho$ is again given by
(\ref{eq:optrho}). \hfill$\square$



\section{Energy Efficiency in the Low-Power Regime}
In this section, we investigate the spectral efficiency--bit energy
tradeoff as the average power $\Pb$ diminishes. We assume that the
bandwidth allocated to the channel is fixed.
With the optimal value of $\rho$ given in Theorem \ref{theo:optrho},
we can now express the normalized effective capacity as
\begin{align}\label{eq:Reimperf}
\R_E(\tsnr,\theta)&=\max_{r\geq0 }- \frac{1}{\theta
TB}\log_e\bigg(1-P\left(|w|^2>\frac{2^{\frac{rT}{TB-1}}-1}{\tsnrefo}\right)\nonumber\\
&\phantom{\max_{r\geq0}- \frac{1}{\theta TB}}\times(1-e^{-\theta
Tr})\bigg)
\end{align}
where \begin{equation}\label{eq:trainsnrefrev}
\tsnrefo=\frac{\phi(\tsnr)\tsnr^2}{\psi(\tsnr)\tsnr+TB-1},
\end{equation}
and
\begin{align}
\phi(\tsnr)=\rhoo(1-\rhoo)\gamma^2T^2B^2,\\
 \text{ and }
\psi(\tsnr)=(1+(TB-2)\rhoo)\gamma TB.
\end{align}
Note that $\tsnr=\Pb/(N_0 B)$ vanishes with decreasing $\Pb$.
We obtain the following result on the bit energy requirement in the low-power regime as $\Pb$ diminishes.
\begin{Lem} \label{theo:imperfect}
In the low-power regime, the bit energy increases
without bound as the average power $\Pb$ and hence $\tsnr$ vanishes,
i.e.,
\begin{gather}
\frac{E_b}{N_0}\bigg|_{\R = 0} = \lim_{\tsnr \to 0} \frac{E_b}{N_0}
= \lim_{\tsnr \to 0} \frac{\tsnr}{\R_E(\tsnr)} =
\frac{1}{\dot{\R_E}(0)} = \infty.
\end{gather}
\end{Lem}
%

This result shows us that operation at very low power levels is extremely energy inefficient and should be avoided regardless of the value of $\theta$. Note that the power allocated for training, $\e_t = \rho \Pb T$, decreases with decreasing $\Pb$. Hence, our ability to estimate the channel is hindered in the low-power regime while, as mentioned before, the system operates as if the channel estimate were perfect. This discrepancy leads to the inefficiency seen as $\Pb$ approaches zero.

\begin{figure}
\begin{center}
\includegraphics[width=\figsize\textwidth]{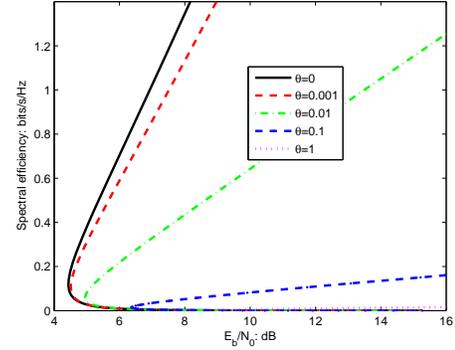}
\caption{Spectral efficiency vs. $E_b/N_0$ in the Rayleigh channel with $E\{|h|^2\} = \gamma = 1$.
$B=10^5$.}\label{fig:6}
\end{center}
\end{figure}

\begin{figure}
\begin{center}
\includegraphics[width=\figsize\textwidth]{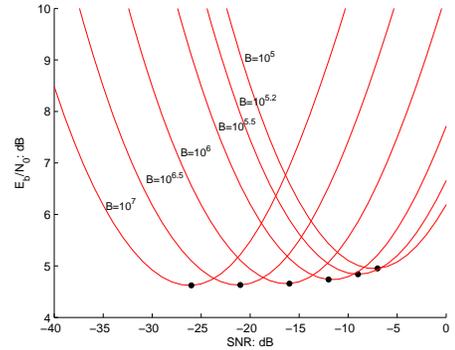}
\caption{$E_b/N_0$ vs. $\tsnr$ in the Rayleigh channel.
$E\{|h|^2\} = \gamma = 1$, $\theta$=0.01.}\label{fig:ebsnr}
\end{center}
\end{figure}

Fig. \ref{fig:6} plots the spectral efficiency vs. bit energy for $\theta=\{1,0.1,0.01,0.001\}$ when $B=10^5$ Hz. As predicted by the result of Theorem \ref{theo:imperfect}, the bit energy increases without bound in all cases as the spectral efficiency $\R_E \to 0$. Consequently, the minimum bit energy is achieved at a nonzero spectral efficiency below which one should avoid operating as it only increases the energy requirements. Another observation is that the minimum bit energy increases as $\theta$ increases and hence as the statistical queueing constraints become more stringent. At higher spectral efficiencies, we again note the increased energy requirements with increasing $\theta$.
In Fig. \ref{fig:ebsnr}, we plot
$\frac{E_b}{N_0}$ as a function of  $\tsnr$ for different bandwidth levels assuming $\theta = 0.01$.
Similarly, we observe that the minimum bit energy is attained at a
nonzero $\tsnr$ value below which $\frac{E_b}{N_0}$ requirements
start increasing. Furthermore, we see that as the bandwidth
increases, the minimum bit energy tends to decrease and is achieved
at a lower $\tsnr$ level.


\section{Energy Efficiency in the Wideband Regime}

In this section, we consider the wideband regime in which the
bandwidth is large. We assume that the average power $\Pb$ is kept
constant. Note that as the bandwidth $B$ increases, $\tsnr =
\frac{\Pb}{N_0B}$ approaches zero and we again operate in the low-$\tsnr$
regime. We denote $\zeta=1/B$. Note that as $B\to \infty$, we have $\zeta\to0$.
With this notation, we can express the normalized effective capacity
as
\begin{equation}
\R_E(\tsnr)=-\frac{\zeta}{\theta
T}\log_e\Big(1-P\{|w|^2>\alphao\}\big(1-e^{-\theta T\ro}\big)\Big).
\end{equation}

The following result provides the expressions for the bit energy at
zero spectral efficiency (i.e., as $B \to \infty$) and the wideband
slope, and characterize the spectral efficiency-bit energy tradeoff
in the wideband regime.
\begin{Lem} \label{theo:wideband}
In the wideband regime, the minimum bit energy and wideband slope
are given by
\begin{gather}
\frac{E_b}{N_0}_{\tmin}=\frac{-\delta\log_e2}{\log_e\xi} \quad \text{and} \label{eq:ebminwb}\\
\mathcal{S}_0=\frac{\xi\log_e^2\xi \log_e2}{\theta T\sta
(1-\xi)\left(\frac{1}{T}\left(\sqrt{1+\frac{\gamma \Pb
T}{N_0}}-1\right)+\frac{\varphi\sta}{2}\right)},\label{eq:s0wb}
\end{gather}
respectively, where $\delta=\frac{\theta T\Pb}{N_0\log_e2}$,
$\xi=1-e^{-\sta}(1-e^{-\frac{\theta T\varphi\sta}{\log_e2}})$, and
$\varphi=\frac{\gamma \Pb}{N_0}\left(\sqrt{1+\frac{N_0}{\gamma \Pb
T}}-\sqrt{\frac{N_0}{\gamma \Pb T}}\right)^2$. $\sta$ is defined as
$\sta=\lim_{\zeta\rightarrow0}\alphao$ and $\sta$ satisfies
\begin{equation} \label{eq:stacondwideband}
\sta=\frac{\log_e2}{\theta T\varphi}\log_e\left(1+\frac{\theta
T\varphi}{\log_e2}\right).
\end{equation}
\end{Lem}

We note that the minimum bit energy in the wideband regime is achieved as $B \to \infty$ and hence as $\tsnr \to 0$. This is in stark contrast to the results in the low-power regime in which the bit energy requirements grow without bound as $\Pb$ and $\tsnr$ vanishes. In the model we have considered, the difference in the performance can be attributed to the fact that increase in the bandwidth, while lowering symbol power $\e_s=\frac{(1-\rho)\Pb
T}{TB-1}$, does not necessarily decrease the training power $\e_t=\rho \Pb T$.

\begin{figure}
\begin{center}
\includegraphics[width=\figsize\textwidth]{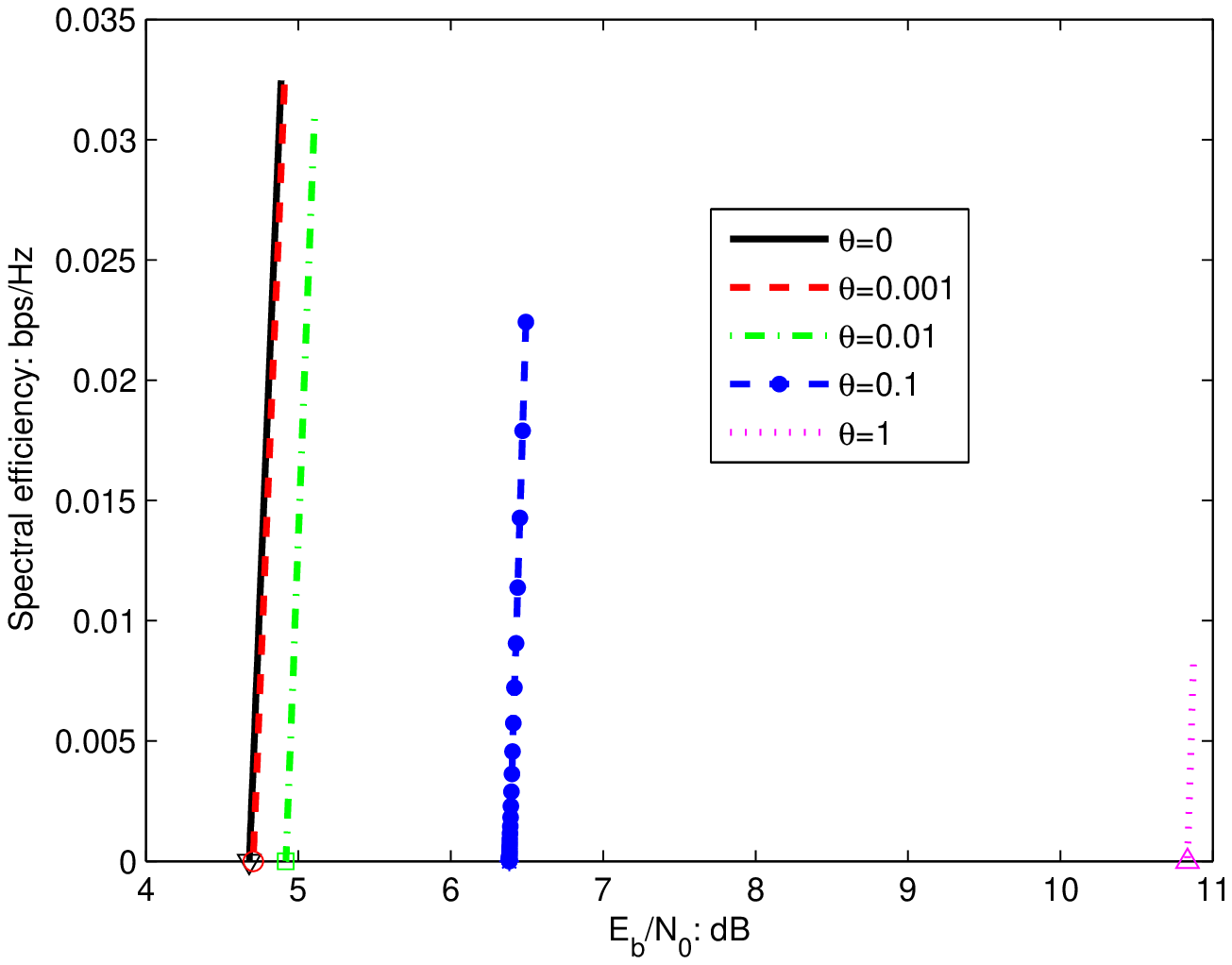}
\caption{Spectral efficiency vs. $E_b/N_0$ in the Rayleigh channel with $E\{|h|^2\} = \gamma = 1$.
$\Pb/N_0=10^4$.}\label{fig:trainwb}
\end{center}
\end{figure}

Fig. \ref{fig:trainwb} plots the spectral efficiency--bit energy curve in
the Rayleigh channel for different $\theta$ values. In the
figure, we assume that $\Pb/N_0=10^4$. As predicted, the minimum bit energies are
obtained as $\tsnr$ and hence the spectral efficiency approach zero.
$\frac{E_b}{N_0}_{\tmin}$ are
computed to be equal to $\{4.6776,4.7029,4.9177,6.3828,10.8333\}$ dB for $\theta=\{0,0.001,0.01,0.1,1\}$, respectively.
Moreover, the wideband slopes are
$\mathcal{S}_0=\{0.4720,0.4749,0.4978,0.6151,0.6061\}$ for the same set of $\theta$ values. As can also be seen in the result of Theorem \ref{theo:wideband}, the minimum bit energy and wideband slope in general depend on $\theta$. In Fig. \ref{fig:trainwb}, we note that the bit energy requirements (including the minimum bit energy) increase with increasing $\theta$, illustrating the energy costs of stringent queueing constraints.

\begin{figure}
\begin{center}
\includegraphics[width=\figsize\textwidth]{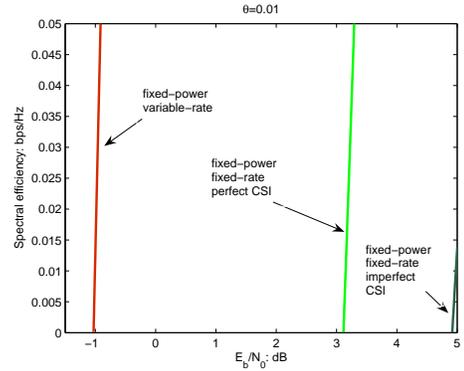}
\caption{Comparison of spectral efficiency; $\Pb/N_0=10^4$,
$\theta=0.01$, and $E\{|h|^2\} = \gamma = 1$.}\label{fig:compair}
\end{center}
\end{figure}

In this paper, we have considered fixed-rate/fixed-power transmissions over imperfectly-known channels. In Fig. \ref{fig:compair}, we compare the performance of this system with those in which the channel is perfectly-known and fixed- or variable-rate transmission is employed. The latter models have been studied in \cite{deli} and \cite{fixed}. This figure demonstrates the energy costs of not knowing the channel and sending the information at fixed-rate.

%

\end{document}